\documentclass[a4paper,11pt]{article}
\usepackage[dvips]{graphicx}
\usepackage{amssymb}
\usepackage{amsmath}
\usepackage{cite}

\begin{document}

\title{$\Lambda$CDM Bounce Cosmology without $\Lambda$CDM: the case of modified gravity}
\author{
S.D. Odintsov,$^{1,2,3}$\,\thanks{odintsov@ieec.uab.es} V.K. Oikonomou,$^{4,5,6}$\,\thanks{v.k.oikonomou1979@gmail.com} \\ \\
$^{1)}$Institut de Ciencies de lEspai (IEEC-CSIC),\\
Campus UAB, Carrer de Can Magrans, s/n\\
08193 Cerdanyola del Valles, Barcelona, Spain\\\\
$^{2)}$ ICREA, Barcelona, Spain \\\\
$^{3)}$ Institute of Physics, Kazan Federal University,\\
Kremlevskaya Street 18,
Kazan 420008, Russia \\\\
$^{4)}$ Department of Theoretical Physics, Aristotle University of Thessaloniki,\\
54124 Thessaloniki, Greece \\\\
$^{5)}$ National Research Tomsk State University, \\
Tomsk, 634050 \\\\
$^{6)}$ Tomsk State Pedagogical University,\\
Tomsk, 634061 Russia \\\\
}

\maketitle

\begin{abstract}
We provide an $F(R)$ gravity description of a $\Lambda$CDM bouncing model, without the need for matter fluids or for cosmological constant. As we explicitly demonstrate, the two cosmological eras that constitute the $\Lambda$CDM bouncing model, can be generated by $F(R)$ gravity which can lead to accelerating cosmologies. The resulting F(R) gravity has Einstein frame inflationary properties that have concordance to the latest Planck observational data. Both the $F(R)$ gravity stability properties are thoroughly investigated and also, the gravitational particle production, a feature necessary for the viability of the $\Lambda$CDM bounce scenario, is also addressed. As we will show, the $\Lambda$CDM bounce model can be successfully described by pure $F(R)$ gravity, with appealing phenomenological attributes, which we extensively discuss.
\end{abstract}


\section*{Introduction}

Recent observational data regarding the Cosmic Microwave Background radiation \cite{wmap,planck} indicate that the scalar perturbations in the early universe were nearly scale invariant, and therefore scale invariance is rendered a fundamental requirement that every cosmological model has to predict to some extent, in order to be considered viable. In most cases, scale invariance is achieved by using scalar fields, the perturbations of which \cite{mukhanov} are scale invariant. Still, no scalar fields have been observed in nature, apart from the Higgs probably, therefore a scenario that avoids scalar fields to some extent can be considered appealing from a physical point of view. One such cosmological scenario was developed by Cai and Ewing \cite{lcdmbounce}, in which case scale invariance is achieved by using only ordinary cold dark matter, radiation and a positive cosmological constant. Particularly, the model describes a bouncing Universe, in which case the Big Bang is replaced by a bounce and therefore the Universe is free from the initial singularity. Bounce cosmologies are known to be alternative scenarios to the standard inflationary cosmology \cite{mukhanov}. For an important stream of reviews and papers on bouncing cosmologies, see \cite{bounce1,bounce2,bounce3,bounce5}. The model developed in \cite{lcdmbounce}, to which we shall refer to as the $\Lambda$CDM bounce model, makes the important assumption that the equation of state, which describes the perfect matter fluids, changes discontinuously and as a consequence of this, the dynamical evolution of the Universe is divided in two cosmological eras. The first era is dynamically governed by radiation with the assumption that quantum gravity effects are taken into account during this era. Particularly, the dynamical evolution is described by loop quantum cosmology (LQC) \cite{LQC1,LQC2,LQC3}, a feature that significantly changes the standard cosmological evolution. The second era is governed by cold dark matter and a positive cosmological constant with the change between the two eras being continuous, with regards to the scale factor. However, it is expected that quantum gravity effects should play an important if not defining role at that universe epoch. Hence, the natural question which appears in relation with above scenario is: can it be realized within some effective gravity model? It is naturally to expect that if such universe can be realized within effective gravity then matter should play minor (if any ) role in its occurrence.

It is one of our main purposes in this paper to provide a pure $F(R)$ gravity description of the $\Lambda$CDM bounce scenario, with pure indicating that no matter fields are going to be used. In addition, we aim to study the stability of the $F(R)$ gravity we shall reconstruct. The $F(R)$ gravity are known to provide consistent theoretical descriptions for cosmological scenarios, which ordinary Einstein-Hilbert gravity fails to describe. For informative reviews on this vast research stream see \cite{reviews1} and for important papers consult Refs. \cite{reviews1,importantpapers1,importantpapers2,importantpapers3,importantpapers4,recontechniques,recontechniques1,recon3,sergeinojirimodel,sergeibabmba,sergeibounce} and references therein.

It is worth to mention that one of the successes of $F(R)$ theories is the consistent description of the late-time acceleration era, with the dark energy finding an appealing and self-consistent geometrical explanation. For alternative theories to $F(R)$ gravity that also provide a theoretically consistent description of dark energy, see for example \cite{capo,capo1,peebles,faraonquin,tsujiintjd,wetterich}.

Moreover, it is known that modified gravity may successfully realize inflation consistent with Planck data (see Refs. \cite{starobinsky,sergeistarobinsky}) or even the unification of inflation with Dark Energy epoch (see Ref. \cite{sergeinojirimodel} and the reviews \cite{reviews1}).

In order to reconcile which $F(R)$ gravity govern the $\Lambda$CDM bounce, we shall make use of two quite well known reconstruction techniques developed in Refs. \cite{importantpapers3} and \cite{recontechniques}. For alternative reconstruction techniques to the ones we shall use, see for example \cite{recontechniques1,recon3}. In addition, for similar studies for other bounces see \cite{sergeibounce,sergeioik,oikonomoubounce}. The reason we shall use to different reconstruction techniques is mainly traced in the particular form of the radiation era of the $\Lambda$CDM bounce. For this case we shall use the technique of \cite{recontechniques} which yields more reliable results in comparison to the technique of \cite{importantpapers3}, as we explain in detail in a future work. With regards to the $\Lambda$CDM era described by matter and cosmological constant, we shall use the reconstruction technique developed in \cite{importantpapers3}, since it yields exact analytical results, without any approximations. In the case of the radiation $\Lambda$CDM bounce era, it is not possible to find an explicit analytic solution, so we investigate this case in the large curvature limit, most relevant to the LQC era which is governed by large curvature quantum operators. As we shall demonstrate, both the eras can be described by $F(R)$ gravity that generate accelerating cosmologies. We also study the stability of our solutions and we investigate in which case instabilities can occur for our solutions. Interestingly enough, one of the two eras is described by an $F(R)$ gravity which when studied in the Einstein frame, can be compatible with the latest Planck data. We study in detail the cases in which this compatibility can be achieved. In addition to these, we give a brief account of the holonomy corrected $F(R)$ gravity \cite{mbouncersquarefr} corresponding to the matter era of the $\Lambda$CDM bounce and we study how the dynamical evolution of the Universe is described in this context. Finally, we address the issue of gravitational particle production, which is a feature that can render the bounce asymmetric, with the latter being favored by observational data.

This paper is organized as follows: In section 1, after providing a brief description of the $\Lambda$CDM bounce and the two eras that it consists of, we make use of the reconstruction techniques to investigate which $F(R)$ gravity generate such a cosmological evolution. In addition, we study the Einstein frame properties of the $F(R)$ gravity that corresponds to the matter era of the $\Lambda$CDM bounce. For the same $F(R)$ gravity we briefly study it's holonomy corrected form in the Einstein frame. The stability of our solutions is thoroughly examined in section 2, while the gravitational particle production issue is addressed in section 3. The conclusions along with a brief discussion on our resulting $F(R)$ gravity picture follow at the end of the paper.

\section{$\Lambda$CDM bounce scenario from $F(R)$ Gravity}

\subsection{A Brief $F(R)$ Gravity Review-Conventions}

In order to make the article self contained we review in brief the essential features of $F(R)$ gravity in the Jordan frame, and also describe the geometrical background that we are working on. There are two approaches in $F(R)$ gravity, namely the Palatini formalism \cite{reviews1} and the metric formalism and we shall work in the context of the latter. The spacetime manifold is assumed to be a pseudo-Riemannian one, which is locally described by a Lorentz metric, and specifically the Friedmann-Robertson-Walker metric in the case at hand. In addition, the metric compatible affine connection is the Levi-Civita connection, which is torsion-less and symmetric. With this connection, the corresponding Christoffel symbols are,
\begin{equation}\label{christofell}
\Gamma_{\mu \nu }^k=\frac{1}{2}g^{k\lambda }(\partial_{\mu }g_{\lambda \nu}+\partial_{\nu
}g_{\lambda \mu}-\partial_{\lambda }g_{\mu \nu}),
\end{equation} 
and additionally, the Ricci scalar is equal to,
\begin{equation}\label{ricciscalar}
R=g^{\mu \nu }(\partial_{\lambda }\Gamma_{\mu \nu }^{\lambda}-\partial_{\nu }\Gamma_{\mu \rho
}^{\rho}-\Gamma_{\sigma \nu }^{\sigma}\Gamma_{\mu \lambda }^{\sigma}+\Gamma_{\mu \rho }^{\rho}g^{\mu
\nu}\Gamma_{\mu \nu }^{\sigma}).
\end{equation}
The $F(R)$ theories are described by the following four dimensional action in the Jordan frame,
\begin{equation}\label{action}
\mathcal{S}=\frac{1}{2\kappa^2}\int \mathrm{d}^4x\sqrt{-g}F(R)+S_m(g_{\mu \nu},\Psi_m),
\end{equation}
with $\kappa^2=8\pi G$ and also $S_m$ encompassing all the matter fields present. The metric formalism of $F(R)$ gravity is actually materialized if the metric tensor it self is considered as the main variable, and by varying action (\ref{action}) with respect to the metric tensor $g_{\mu \nu}$, we acquire the following equations of motion
\begin{equation}\label{eqnmotion}
F'(R)R_{\mu \nu}(g)-\frac{1}{2}F(R)g_{\mu \nu}-\nabla_{\mu}\nabla_{\nu}F'(R)+g_{\mu \nu}\square
F'(R)=\kappa^2T_{\mu \nu}^m.
\end{equation} 
The prime in Eq. (\ref{eqnmotion}) denotes differentiation with respect to the argument of the differentiated function, that is, $F'(R)=\partial F(R)/\partial R$ and additionally $T_{\mu \nu}^m$ denotes the energy-momentum tensor of the matter fields. Finally, as we already mentioned, the metric will be assumed to be a flat Friedmann-Lemaitre-Robertson-Walker (FRW hereafter), with the corresponding line element being equal to, 
\begin{equation}\label{metricformfrwhjkh}
\mathrm{d}s^2=-\mathrm{d}t^2+a^2(t)\sum_i\mathrm{d}x_i^2,
\end{equation}
and $a(t)$ being the scale factor. The Ricci scalar corresponding to this line element is,
\begin{equation}\label{ricciscal}
R=6(2H^2+\dot{H}),
\end{equation}
with $H(t)$ denoting the Hubble parameter $H(t)=\dot{a}/a$, and the dot indicates time differentiation.

\subsection{An overview of the $\Lambda$CDM bounce scenario}

The focus in this article is on the cosmological scenario described in detail in Ref. \cite{lcdmbounce} by Cai and Ewing. The scenario itself is quite physically appealing and we shall provide a pure $F(R)$ gravity description of it. Before we proceed to the $F(R)$ gravity reconstruction of the cosmological scenario, it is worth to provide a detailed description of it, since it is of importance to understand the new insights that the $F(R)$ gravity brings along. For details on this scenario, the reader is referred to Ref. \cite{lcdmbounce}. In addition, for similar models with distinct cosmological eras, see \cite{piao}. 

The cosmological model described in \cite{lcdmbounce}, described the dynamics of a flat FRW cosmology with a positive cosmological constant and also with radiation and cold dark matter (CDM hereafter) present. The cosmological evolution was divided in two eras, namely one described by an effective LQC Hamiltonian, in which the minisuperspace wave function is a sharply peaked state, very adequately described by the effective equations of state of LQC, and the other evolution era is described by a cosmological constant plus CDM. 

With regards to the effective LQC cosmological era, it was assumed that radiation dominates this regime, which is considered to be the high curvature regime, in which case the effective equations of motion are given by \cite{LQC1,LQC2,LQC3,mbouncersquarefr,lcdmbounce},
\begin{equation}\label{EFFLQC}
H^2=\frac{8\pi G}{3}\rho \left (1-\frac{\rho}{\rho_c} \right ),
\end{equation}
with $H$ the Hubble rate, $\rho$ the radiation matter fluid energy-density and $\rho_c\sim \rho_{pl}$, the critical energy-density which is of the order of the Planck energy-density. In the radiation-LQC era, the quantum gravity effects are assumed to control the cosmological evolution, as is obvious from the holonomy corrected FRW equation (\ref{EFFLQC}). With regards to the cosmological constant plus CDM era, which in the rest of the paper we shall refer to as matter-CC phase, the cosmological evolution is governed by the cosmological constant and CDM.

In addition, and more importantly, it is assumed that the Universe's evolution is broken into the aforementioned cosmological eras, and that there is a discontinuous transition between these two eras, a fact that will be clearly depicted in the equations of state describing the two eras. As we explicitly demonstrate, this equation of state discontinuity will also be materialized in our pure $F(R)$ gravity description, but we do not include any matter fluids in our description. Particularly, as we shall show, the two $F(R)$ gravity that correspond to the two distinct cosmological eras, are mathematically distinct, but interestingly enough, both produce accelerating cosmologies in the large curvature limit. Specifically, the LQC radiation era is described by an $R+\Lambda$ gravity, and the $\Lambda$CDM era is described by an $R^2$ gravity, in the large curvature regime.

Having assumed a discontinuous evolution of the Universe, in the aforementioned eras, let us briefly describe these two eras, the scale factor of which we shall extensively use in the following sections. The LQC radiation era is governed by radiation, with an equation of state, 
\begin{equation}\label{eqnstate}
\rho (t)=\frac{\rho_c}{a(t)^4},
\end{equation}
and in conjunction with Eq. (\ref{EFFLQC}) we obtain the LQC radiation era scale factor \cite{lcdmbounce},
\begin{equation}\label{lqcradscale}
a(t)=\left ( \frac{32\pi G\rho_c}{3}t^2+1\right )^{1/4}.
\end{equation}
In the earlier epoch, the curvature is lower and quantum gravity effects no longer govern the evolution of the Universe, a process which now is governed by the CDM energy-density and cosmological constant, which we denote $\rho_{CDM}$ and $\Lambda$ respectively. In this case, the FRW equations are given by the following expression,
\begin{equation}\label{frweqns}
H^2=\frac{8\pi G}{3}a(t)^2\left (\rho_{CDM}+\frac{\rho_{\Lambda}}{8\pi G}\right ).
\end{equation}
Denoting the total energy density $\rho_{tot}$, it is obvious that 
\begin{equation}\label{asxeton}
\rho_{tot}=\rho_{CDM}+\rho_{\Lambda}.
\end{equation}
It is also assumed that the equation of state is of the form $P_{tot}=\omega \rho_{tot}$, with $\omega=-\delta$ and $0\leq \delta \leq 1$. The parameter $\delta$ is assumed to vary continuously from the cosmological constant epoch to the CDM matter epoch, but it is also assumed that when a specific era is considered, this is almost constant \cite{lcdmbounce}. In this context, the total energy density for a specific epoch is given by,
\begin{equation}\label{specfic}
\rho_{tot}=\frac{\rho_{eff}}{a^{3(1-\delta)}}.
\end{equation}
with $\rho_{eff}$ a constant related to the scale factor at the radiation matter era transition, which for simplicity we leave it as $\rho_{eff}$. Notice that the value $\delta=1$ corresponds to the cosmological constant epoch ($\omega=-1$), while the value $\delta=0$ corresponds to the CDM epoch ($\omega=0$). The scale factor for the fluid with equation of state $P_{tot}=\omega \rho_{tot}$, as a function of the cosmological time $t$ and of the parameter $\delta$, is equal to,
\begin{equation}\label{lcdmscale}
a(t)=\mathcal{A}\left (t-\gamma \right )^{\frac{2}{3(1-\delta )}},
\end{equation}
where we have set $A$ to be equal to,
\begin{equation}\label{alphaparameter}
\mathcal{A}=\left ( \sqrt{\frac{2\pi G\rho_{eff}}{3}}(1-3\delta)\right )^{\frac{2}{1-3\delta }}\left (\frac{3-3\delta}{1-3\delta}\right )^{\frac{1-3\delta}{3-3\delta}}.
\end{equation}
The parameter $\gamma$ appearing in Eq. (\ref{lcdmscale}) is again related to the transition time from radiation to matter era, but it's exact definition will play no important role in our analysis, for details see \cite{lcdmbounce}. In the following sections, we shall thoroughly investigate which pure $F(R)$ gravity can generate the cosmologies described by Eqs. (\ref{lqcradscale}) and (\ref{lcdmscale}). Special emphasis shall be given in the transition epoch between the two eras, and specifically in the high curvature regime, when the LQC era ends and the $\Lambda$CDM era starts (and in particular the cosmological constant era). This era is of particular importance with regards to the $F(R)$ gravity we shall find, since it can describe the inflationary era in the Einstein frame. We shall study in detail the resulting picture in a following section.

\subsection{$\Lambda$CDM bounce from $F(R)$ Gravity: The matter-cosmological constant phase}

We start our analysis with the reproduction of the matter-CC phase by a pure $F(R)$ gravity. Our strategy is to find which pure $F(R)$ gravity can produce a cosmological evolution with a scale factor equal to the one of Eq. (\ref{lcdmscale}). We shall use the reconstruction technique of Ref. \cite{importantpapers3}, which makes use of the e-fold number $N$. This is most appropriate for the case at hand, since this technique leads to differential equations which can be solved analytically. There is an equally useful reconstruction technique developed in \cite{recontechniques}, which makes use of an auxiliary field. This technique yields approximate results to the large and small curvatures limits, but since we can have exact analytic results with the technique developed in \cite{importantpapers3}, we use the latter. In a following section we shall have a small discussion on the issue of choosing the best technique, but let us mention that the two techniques yield the same results, if these are used properly.

The Hubble rate corresponding to the scale factor (\ref{lcdmscale}) is equal to,
\begin{equation}\label{hubble1}
H(t)=\frac{2}{3 (1-\delta)(t-\gamma)},
\end{equation}
and recall that $0\leq \delta \leq 1$. The first FRW equation is written in the following way,
\begin{equation}\label{frwf1}
-18\left ( 4H(t)^2\dot{H}(t)+H(t)\ddot{H}(t)\right )F''(R)+3\left (H^2(t)+\dot{H}(t) \right )F'(R)-\frac{F(R)}{2}=0,
\end{equation}
with $F'(R)=\frac{\mathrm{d}F(R)}{\mathrm{d}R}$ and the Ricci scalar $R$ is given in Eq. (\ref{ricciscal}). The e-folding number $N$ is related to the scale factor in the following way,
\begin{equation}\label{efoldpoar}
e^{-N}=\frac{a_0}{a},
\end{equation} 
and by using this variable $N$, the first FRW equation can be expressed in terms of the e-fold parameter $N$ as follows,
\begin{align}\label{newfrw1new1}
& -18\left ( 4H^3(N)H'(N)+H^2(N)(H')^2+H^3(N)H''(N) \right )F''(R)
\\ \notag & +3\left (H^2(N)+H(N)H'(N) \right )F'(R)-\frac{F(R)}{2}=0.
\end{align}
In the relation above, the derivatives are defined with respect to the new variable $N$, that is, $H'=\mathrm{d}H/\mathrm{d}N$ and $H''=\mathrm{d}^2H/\mathrm{d}N^2$ and the same convention holds true in the rest of this section. Introducing the function $G(N)=H^2(N)$ in equation (\ref{newfrw1new1}), the latter can be recast as follows,
 \begin{align}\label{newfrw1modfrom}
& -9G(N(R))\left ( 4G'(N(R))+G''(N(R)) \right )F''(R)
\\ \notag & +\left (3G(N)+\frac{3}{2}G'(N(R)) \right )F'(R)-\frac{F(R)}{2}=0,
\end{align}
with $G'(N)=\mathrm{d}G(N)/\mathrm{d}N$ and $G''(N)=\mathrm{d}^2G(N)/\mathrm{d}N^2$. A crucial point is the fact that
\begin{equation}\label{riccinrelat}
R=3G'(N)+12G(N),
\end{equation}
and by using this in conjunction with (\ref{efoldpoar}), we end up to a differential equation which will provide us with the reconstructed $F(R)$ gravity which produces the cosmology (\ref{lcdmscale}). Indeed, by writing the Hubble rate as a function of the scale factor,
\begin{equation}\label{hpscf}
H=\frac{2}{3 (1-\delta)}a^{-\frac{3 (1-\delta)}{2}} ,
\end{equation}
and by making use of (\ref{efoldpoar}) and also recalling that $G(N)=H^2(N)$, we have,
\begin{equation}\label{gnfunction}
G(N)=A e^{-3 (1-\delta)N},
\end{equation}
where $A$ is equal to $A=\frac{4}{9 (1-\delta)^2}a_0^{-3 (1-\delta)}$. Thereby, by using Eqs. (\ref{riccinrelat}) and (\ref{gnfunction}), the e-fold number can be expressed as a function of the Ricci scalar,
\begin{equation}\label{efoldr}
N=-\frac{1}{3 (1-\delta)}\ln \left ( \frac{R}{B}\right ),
\end{equation}
with $B=-9A (1-\delta)+12A$, and so the differential equation of Eq. (\ref{newfrw1modfrom}) takes the form, 
\begin{align}\label{bigdiffgeneral1}
&-\frac{9A^23 (1-\delta)(3 (1-\delta)-4)}{B^2}R^2\frac{\mathrm{d}^2F(R)}{\mathrm{d}R^2}
-\frac{3A(3 (1-\delta)-2)}{2B}R\frac{\mathrm{d}F(R)}{\mathrm{d}R}-\frac{F(R)}{2}=0.
\end{align}
Notice that we assumed that no matter fluids are present so that the differential equation (\ref{bigdiffgeneral1}) yields the pure $F(R)$ gravity that generates (\ref{lcdmscale}). In order to simplify the notation of the equations to follow, we introduce the parameters $a_1$ and $a_2$, which are defined to be,
\begin{equation}\label{apara1a2}
a_1=-\frac{9A^23 (1-\delta)(3 (1-\delta)-4)}{B^2},{\,}{\,}a_2=-\frac{3A(3 (1-\delta)-2)}{2B},
\end{equation}
The differential equation (\ref{bigdiffgeneral1}) is the homogeneous Euler second order differential equation, the solutions of which we denote by $f_1(R)$ and $f_2(R)$. These are equal to,
\begin{equation}\label{dgfressss}
f_1(R)=R^{\rho_1},{\,}{\,}{\,}f_2(R)=R^{-\rho_2},
\end{equation}
where the parameters $\rho_1$ and $\rho_2$ are given below,
\begin{equation}\label{rho12}
\rho_1=\frac{-(a_1-a_1)+\sqrt{(a_2-a_1)^2+2a_1}}{2a_1},{\,}{\,}{\,}\rho_2=\frac{-(a_1-a_1)-\sqrt{(a_2-a_1)^2+2a_1}}{2a_1},
\end{equation}
and with $a_1,a_2$ given in (\ref{apara1a2}). Hence, the pure $F(R)$ that generates $\Lambda$CDM bounce solution of relation (\ref{lcdmscale}) is the following,
\begin{equation}\label{frgenerlargetssss}
F(R)=c_1R^{\rho_1}+c_2R^{\rho_2}
\end{equation}
with $c_1,c_2$ free parameters. It is worth providing the exact relation of the variables $\rho_1$ and $\rho_2$ as functions of $\delta$, which is,
\begin{equation}\label{rho12deltanew}
\rho_1=\frac{2}{3-\frac{10}{1+3 \delta }+\sqrt{\frac{73-78 \delta +9 \delta ^2}{(1+3 \delta )^2}}},{\,}{\,}{\,}\rho_2=\frac{1}{\frac{3}{2}-\frac{5}{1+3 \delta }-\frac{1}{2} \sqrt{\frac{73-78 \delta +9 \delta ^2}{(1+3 \delta )^2}}}.
\end{equation}
The resulting $F(R)$ gravity varies as a function of the parameter $\delta$ that determines the equation of state during the matter-CC phase of the $\Lambda$CDM bounce. We shall investigate the form of the $F(R)$ gravity for the limiting values of $\delta$ and also examine the large curvature limit which is very essential with regards to the other evolution era of the scenario we study, namely the LQC-radiation era.

The most interesting case from a physical point of view is when $\delta \rightarrow 1$, which case corresponds to an equation of state that describes a cosmological constant. For $\delta \rightarrow 1$, the parameters $\rho_1$ and $\rho_2$ behave as follows,
\begin{equation}\label{behrhoinf2}
\rho_1=2,{\,}{\,}{\,}\lim_{\delta\rightarrow 1}\rho_2 \rightarrow -\infty,
\end{equation}
and therefore for $\delta \rightarrow 1$ and in the large curvature regime, the $F(R)$ gravity behaves as,
\begin{equation}\label{largecurvdeltaone}
F(R)\sim c_1 R^2,
\end{equation}
since the term $R^{\rho_2}$ is almost zero as $R$ tends to infinity. We have therefore ended up to a very physical appealing scenario in which the large curvature $F(R)$ gravity which describes the matter-CC phase is an $R^2$ gravity, when $\delta=1$. The $R^2$ gravity is known to have quite interesting inflation properties \cite{starobinsky,sergeistarobinsky}. This result is of great importance and we will discuss it in a later section, when we also have the $F(R)$ gravity which generates the LQC-radiation era. From the form of the parameters $\rho_1$ and $\rho_2$ as functions of $\delta$, it is obvious that in the large curvature regime only the $\rho_1$ term dominates, so the $F(R)$ gravity for a general value of $\delta$ in the large curvature regime is,
\begin{equation}\label{largecurvdeltaonelarge}
F(R)\sim c_1 R^{\rho_1}.
\end{equation}
For $\delta=0$, which corresponds to the pure matter domination era, the parameter $\rho_1$ is equal to,
\begin{equation}\label{unimportant}
\rho_1=\frac{1}{12} \left(7+\sqrt{73}\right),
\end{equation}
for which case, nothing interesting occurs. However, having an $R^2$ gravity describing the large curvature regime, it is worth investigating if there are values of $\delta$ for which we can have some overlap with the Planck observational data \cite{planck}. In the next section we shall investigate the properties of the large curvature $F(R)$ gravity (\ref{largecurvdeltaonelarge}) in the Einstein frame and as we explicitly demonstrate, there are values of $\delta$ for which we can achieve exact compatibility with the latest Planck data \cite{planck}.

\subsubsection{Einstein frame inflation study of the $\Lambda$CDM bounce $F(R)$ gravity}

As we evince in this section, the $F(R)$ gravity of Eq. (\ref{frgenerlargetssss}) in the large curvature limit, when conformally transformed in the Einstein frame, can yield results that overlap to a great extent with the Planck data. To start with, the $F(R)$ gravity (\ref{frgenerlargetssss}), in the large $R$ limit gets simplified and is given by (\ref{largecurvdeltaonelarge}). A detailed analysis on the Einstein frame inflation properties of $F(R)$ gravity was performed in \cite{sergeistarobinsky} and we adopt the notation of this reference. In order to make contact with \cite{sergeistarobinsky}, we identify our parameter $\rho_1$ with the following expression,
\begin{equation}\label{identification}
\rho_1=\frac{n+2}{n+1},
\end{equation}
and also we set $c_1$ to be equal to,
\begin{equation}\label{constantcq1}
c_1=\gamma \left ( \frac{n+1}{n+2}\right ) \left ( \frac{1}{4(n+2)}\right )^{1/(n+1)}.
\end{equation}
since $c_1$ is a free variable. Then, the $F(R)$ gravity in the large curvature limit reads,
\begin{equation}\label{largefr}
F(R)\simeq \gamma \left ( \frac{n+1}{n+2}\right ) \left ( \frac{1}{4(n+2)}\right )^{1/(n+1)}\left ( \frac{R}{\gamma}\right )^{\frac{n+2}{n+1}}.
\end{equation}
Before continuing and making contact with observational data we need to discuss the theoretical framework of the Einstein frame. As we discussed in the previous sections, the $F(R)$ gravity (\ref{largefr}) generates the matter-CC phase FRW cosmology with the scale factor of Eq. (\ref{lcdmscale}), so this does not generate inflation in the Jordan frame. Our aim is to study the Einstein frame inflation properties of (\ref{largefr}), so we assume that in the Jordan frame, the metric is an appropriately chosen one, so that when conformally transformed in the Einstein frame, it yields an inflation generating de Sitter or at least quasi-de Sitter metric. Then by conformally transforming the Jordan frame theory using standard techniques \cite{reviews1,sergeistarobinsky}, we obtain the following scalar potential, 
\begin{equation}\label{scalarpotentialeinsteinframe}
V(\sigma )=-\frac{\gamma (n+2)}{k^2}\left (1-e^{\sqrt{\frac{2}{3}k^2} \sigma }\right )+\frac{\gamma}{k^2}e^{-n\sqrt{\frac{2}{3}k^2} \sigma }.
\end{equation}
The slow-roll parameters are defined as,
\begin{equation}\label{slowrollinflpotent}
\epsilon =\frac{1}{2k^2}\left (\frac{V'(\sigma)}{V(\sigma)}\right ),{\,}{\,}{\,}\eta =\frac{1}{k^2}\left ( \frac{V''(\sigma )}{V(\sigma )} \right ),
\end{equation}
during the slow-roll inflation, and for the potential (\ref{scalarpotentialeinsteinframe}), these are equal to,
\begin{align}\label{slowrollinflpotent1}
& \epsilon \simeq \frac{\left ( n-(n+2)e^{(n+1)\sqrt{\frac{2}{3}k^2} \sigma }\right )^2}{(n+2)e^{n\sqrt{\frac{2}{3}k^2} \sigma }-(n+2)e^{(n+1)\sqrt{\frac{2}{3}k^2} \sigma }-1}
\\ \notag & \eta =\frac{2}{3}\frac{n^2+(n+2)e^{(n+1)\sqrt{\frac{2}{3}k^2} \sigma }}{1+(n+2)e^{(n+1)\sqrt{\frac{2}{3}k^2} \sigma }-(n+2)e^{n\sqrt{\frac{2}{3}k^2} \sigma }}.
\end{align}
These slow parameters at the limit $\sigma\rightarrow -\infty $, become approximately equal to,
\begin{equation}\label{apprxslowroll}
\epsilon \simeq \frac{n^2}{3},{\,}{\,}{\,}\eta \simeq 2\frac{n^2}{3}.
\end{equation}
So eventually the primordial power spectrum and the spectral indexes of inflation are equal to,
\begin{equation}\label{primordandinflationindexes}
\Delta_R^2\simeq \frac{k^2\gamma e^{\frac{2}{3}Nn^2}}{8\pi^2n^2},{\,}{\,}{\,}n_s\simeq 1-2\frac{n^2}{3},{\,}{\,}{\,}r\simeq 16\frac{n^2}{3},
\end{equation}
where we omitted corrections of the order $e^{-\frac{2n^2N}{3}}$. The latest observational data constrain the spectral indexes as follows,
\begin{equation}\label{spectrconstraints}
n_s=0.9603\pm 0.0073,{\,}{\,}{\,}r<0.11,
\end{equation} 
so the spectral indexes of Eq. (\ref{primordandinflationindexes}) are  consistent with the observational data when $n\simeq 0.2$ and $n\simeq 0.1$. Recalling relation (\ref{identification}) and since the parameter $\rho_1$ is related to $\delta$ as follows,
\begin{equation}\label{sdgegdelta}
\rho_1=\frac {2} {3 - \frac {10} {1 + 
      3\delta} + \sqrt {\frac {73 - 78\delta + 
        9\delta^2} {(1 + 3\delta)^2}}},
\end{equation}
the values $n\simeq 0.2$ and $n\simeq 0.1$ correspond to $\delta\simeq 0.95$ and $\delta\simeq 0.97$. Of course these values are allowed, since $0\leq \delta \leq 1$. In the following we shall take into account these two values of $\delta$ and the corresponding in each case $F(R)$ gravity, which for $\delta=0.95$ reads,
\begin{equation}\label{dfractyalvalue}
F(R)\simeq c_1R^{2-0.166667},
\end{equation}
while for $\delta=0.97$ is equal to,
\begin{equation}\label{dfractyalvalue1}
F(R)\simeq c_1 R^{2-0.111111}.
\end{equation}

\subsubsection{Holonomy Corrected $F(R)$ gravity and the $\Lambda$CDM bounce: An alternative approach to inflation}

As we already mentioned, the $F(R)$ gravity that describes the large curvature regime of the matter-CC phase is of the form $\sim R^{\rho_1}$, and during this regime, the quantum gravity effects are disregarded. According to scenario of Cai and Ewing, the quantum gravity effects effectively modify the LQC-radiation phase. Before proceeding to the $F(R)$ description of this cosmological phase, we shall study in this section how holonomy corrections may modify the cosmological evolution of the $F(R)$ gravity given in Eq. (\ref{largecurvdeltaonelarge}). As we shall show, the holonomy corrected $F(R)$ gravity in the Einstein frame can successfully describe a bouncing cosmology itself, without the need for another cosmological era. We intend to give a brief qualitative analysis of the holonomy corrected $\sim R^{\rho_1}$ gravity in the Einstein frame, but we intend to address the problem more concretely and in the Jordan frame in a future publication.

The holonomy corrected $F(R)$ gravity in the Einstein frame were studied in \cite{mbouncersquarefr}. Introducing the Einstein frame holonomy corrections, the FRW equation reads,
\begin{equation}\label{einframeholcorr}
\tilde{H}^2=\frac{1}{3}\tilde{\rho}\left ( 1-\frac{\tilde{\rho}}{\tilde{\rho_c}}\right )
\end{equation}
with $\tilde{\rho}_c$ the Einstein frame critical density. Equation (\ref{einframeholcorr}) describes an ellipse in the $(\tilde{H},\tilde{\rho })$ plane, and the Universe's evolution is quite simple to describe. Particularly, the universe is moving clockwise starting from a contracting phase and proceeding to an expanding phase. Both phases begin and end at the same critical point $(0,0)$ and the system bounces off only once at the point $(0,\tilde{\rho })$. Using the scalar potential in the Einstein frame (\ref{scalarpotentialeinsteinframe}), the Einstein frame scalar field evolution is equal to,
\begin{equation}\label{scalrevol}
\ddot{\sigma}+3\tilde{H}\dot{\sigma }+\frac{\partial V(\sigma)}{\partial{ \sigma}}=0,
\end{equation}
Performing the transformation $\sqrt{\frac{2k^2}{3}}\psi =\ln \sigma$, the Einstein frame evolution of the new scalar field $\psi$ is governed by the following differential equation,
\begin{align}\label{psievol}
& \ddot{\psi}\psi+3\tilde{H}\dot{\psi }\psi+
\\ \notag & +\frac{1}{\sqrt{6}(\rho_1-1)k}\Big{(}(\rho_1-2)(\rho_1 A_4)^{\frac{1}{1-\rho_1}}\psi^{g+1}-2(\rho_1-1)(\rho_1 A_4)^{\frac{\rho_1}{1-\rho_1}}\psi^{g} \Big{)}
\end{align}
with the parameter ''$g$'' set equal to,
\begin{equation}\label{deltan}
g=\frac{2\rho_1-3}{\rho_1-1}
\end{equation}
The bounce produced from the dynamical system (\ref{psievol}) is symmetric, a fact that makes this scenario less appealing, as we shall discuss later on. It is easy to see how a symmetric bounce is generated without solving the differential equation (\ref{psievol}) explicitly, since it possesses some symmetries that make the qualitative analysis quite easy. Indeed, the orbits of the dynamical system (\ref{psievol}), depicted in phase space by $(\dot{\psi},\psi )$, are symmetric around the $\dot{\psi}$, owing to the fact that Eq. (\ref{psievol}) remains invariant under the following transformations,  
\begin{equation}\label{ttransfrom}
\tilde{t}\rightarrow -\tilde{t},{\,}{\,}{\,}\tilde{H}\rightarrow -\tilde{H}
\end{equation}
Therefore, the contracting phase ($\tilde{H}<0$) orbit $(\psi(t),\dot{\psi}(t))$ under the transformation (\ref{ttransfrom}) is transformed to the expanding phase ($\tilde{H}>0$) orbit $(\psi(-t),\dot{\psi}(-t))$ and thus, the bounce is symmetric. In addition, the Einstein frame energy density $\tilde{\rho}$ is equal to,
\begin{equation}\label{energydensityeinframe}
\tilde{\rho}=\frac{\dot{\psi}^2}{2\psi^2}+\frac{1}{2k^2\psi^2}\left ( (\rho_1 A_4)^{\frac{1}{1-\rho_1 }}\psi^{n+1}-A_4(\rho_1 A_4)^{\frac{\rho_1}{1-\rho_1 }}\psi^{n}\right )
\end{equation}
The Hubble parameter $\tilde{H}$ is directly related to the energy-density $\tilde{\rho}$ by the holonomy corrected FRW equation (\ref{einframeholcorr}) and as it is obvious, the Hubble parameter vanishes at the point $(\psi,\dot{\psi})=(\delta,0)$ and in addition at the curve $\tilde{\rho}=\tilde{\rho}_c$. A simple qualitative analysis of this result may be done easily and it goes as follows: The universes evolution begins in the contracting phase with $\tilde{H}<0$ and oscillates around the point $(\delta,0)$, where the oscillations amplitude increases up to the point it reaches the curve $\tilde{\rho}=\tilde{\rho}_c$, at which $\tilde{H}=0$. At this point the universe bounces off and it enters the expanding phase, during which $\tilde{H}>0$. The expansion continues in an oscillating way until the critical point $(\delta,0)$ is reached. 

Before we close this qualitative description we have to note two things. Firstly, the $\tilde{\rho}=\tilde{\rho}_c$ curve is not simple compared to the $R^2$ gravity curve studied in \cite{mbouncersquarefr} and secondly the bounce predicted by this scenario is symmetric. A symmetric bounce however is ruled out, as was also pointed out by Cai and Ewing \cite{lcdmbounce}, so the only way to achieve an asymmetric bounce within this framework is only through particle production enhancement. We shall address this issue in a later section.

\subsection{$\Lambda$CDM bounce from $F(R)$ gravity in the large curvature limit: The radiation phase}

The Hubble parameter corresponding to the scale factor (\ref{lqcradscale}) which describes the LQC-radiation phase, is equal to,
\begin{equation}\label{holcor1}
H(t)=\frac{\frac{32\pi G\rho_ct}{3}}{\frac{64\pi G\rho_ct^2}{3}+2}.
\end{equation}
This Hubble parameter will be our starting point for the reconstruction of the LQC-radiation phase cosmology. At this point we have to make a crucial remark, related to the issue of choosing the most optimal reconstruction method. Particularly, since we are interested in the large curvature regime of the LQC-radiation phase, this corresponds to early times which means small values of the cosmological time. Particularly, as also noted by the authors of Ref. \cite{lcdmbounce}, the fact that for small cosmological times the scale factor is non-zero is a manifestation of quantum gravity effects. Taking the limit of the scale factor (\ref{lqcradscale}) when $t$ tends to zero, we obtain that indeed the scale factor is non-zero and equal to,
\begin{equation}\label{limsclfzero}
a(t)=1.
\end{equation}
This is a crucial observation, since the reconstruction method we used in the previous section, requires some functional dependence of the scale factor with respect to time, so the most optimal method for reconstructing the cosmology described by the Hubble rate (\ref{holcor1}), is the reconstruction method developed in \cite{recontechniques}, which we use in this section. Disregarding any contribution form matter fluids, the first FRW equation reads,
\begin{equation}\label{frwf1}
-18\left ( 4H(t)^2\dot{H}(t)+H(t)\ddot{H}(t)\right )F''(R)+3\left (H^2(t)+\dot{H}(t) \right )F'(R)-\frac{F(R)}{2}=0.
\end{equation}
The reconstruction method of \cite{recontechniques}, makes use of an auxiliary field $\phi$, so that the action (\ref{action}) which describes the pure $F(R)$ gravity, is modified in the following way,
\begin{equation}\label{neweqn123}
S=\int \mathrm{d}^4x\sqrt{-g}\left ( P(\phi )R+Q(\phi ) +\mathcal{L}_{mat} \right ).
\end{equation}
Practically, the final form of the reconstructed $F(R)$ gravity will be given by the functions $P(\phi )$ and $Q(\phi )$, so the focus is to find these solutions explicitly. The absence of a kinetic term for the scalar field in relation (\ref{neweqn123}), renders it an auxiliary time dependent degree of freedom, so upon variation with respect to $\phi$, we obtain, 
\begin{equation}\label{auxiliaryeqns}
P'(\phi )R+Q'(\phi )=0
\end{equation}
where the prime denotes differentiation with respect to $\phi$. Solving this algebraic relation with respect to $\phi$, will yield the function $\phi (R)$ and the $F(R)$ gravity can be found easily by substituting $\phi (R)$ to action (\ref{neweqn123}), so that $F(R)$ is equal to,
\begin{equation}\label{r1}
F(\phi( R))= P (\phi (R))R+Q (\phi (R)).
\end{equation}
Therefore, it is of critical importance to modify the FRW equation in such a way so that we obtain a differential equation that will yield $P(\phi)$ and $Q(\phi )$. This differential equation can be easily obtained by varying equation (\ref{neweqn123}) with respect to the metric tensor. Assuming a flat FRW metric, we obtain,
\begin{align}\label{r2}
& -6H^2P(\phi (t))-Q(\phi (t) )-6H\frac{\mathrm{d}P\left (\phi (t)\right )}{\mathrm{d}t}=0 \\ \notag &
\left ( 4\dot{H}+6H^2 \right ) P(\phi (t))+Q(\phi (t) )+2\frac{\mathrm{d}^2P(\phi (t))}{\mathrm {d}t^2}+\frac{\mathrm{d}P(\phi (t))}{\mathrm{d}t}=0.
\end{align}
By eliminating the function $Q(\phi (t))$ we obtain,
\begin{equation}\label{r3}
2\frac{\mathrm{d}^2P(\phi (t))}{\mathrm {d}t^2}-2H(t) P(\phi (t))+4\dot{H}\frac{\mathrm{d}P(\phi (t))}{\mathrm{d}t}=0.
\end{equation}
As it is explicitly proven in \cite{recontechniques}, the scalar field $\phi$ is considered to be equivalent to the cosmological time $t$, so in the following we assume that $\phi=t$ (see appendix of \cite{recontechniques}). Assuming that the scale factor takes the form,
\begin{equation}\label{r4}
a=a_0e^{g(t)}
\end{equation}
with $a_0$ being a constant, the differential equation (\ref{r3}) can be recast in the following way,
\begin{align}\label{r5}
& 2\frac{\mathrm{d}^2P(\phi (t))}{\mathrm {d}t^2}-2g'(\phi )\frac{\mathrm{d}P(\phi (t))}{\mathrm{d}t}+4g''(\phi ) P(\phi (t))=0.
\end{align}
This differential equation yields $P(\phi )$, and by using the resulting $P(\phi )$, we may get the $Q(\phi )$,
\begin{equation}\label{r5a}
Q(\phi )=-6g'(\phi )^2P(\phi )-6g'(\phi )\frac{\mathrm{d}P(\phi )}{\mathrm{d}\phi }.
\end{equation} 
The Hubble rate (\ref{holcor1}) can be written in the following form,
\begin{equation}\label{r6}
H(t)=\frac{h(t)}{t},
\end{equation}
with $h(t)$ being equal to,
\begin{equation}\label{r7}
h(t)=\frac{h_fqt^2}{1+qt^2},
\end{equation}
and where we defined $h_f$ and $q$ to be equal to,
\begin{equation}\label{r7a}
h_f=\frac{1}{2},{\,}{\,}{\,}q=\frac{32\pi G\rho_c}{3}.
\end{equation}
The function $h(t)$ appearing in Eq. (\ref{r7}) is a slowly varying function of time, a property that will significantly simplify the problem at hand. Indeed, the function $h(t)$ $\forall $ $z$ $\in $ $\mathbb{R}$ satisfies the following relation,
\begin{equation}\label{r10}
\lim_{t\rightarrow \infty}\frac{h(zt)}{h(t)}=1.
\end{equation}
We therefore assume a particular form for the function appearing in relation (\ref{r4}) which is,
\begin{equation}\label{r11}
g(\phi )=h(\phi)\ln \left( \frac{\phi}{\phi_0}\right ),
\end{equation}
with $\phi_0$ some integration constant. Since the function $h(\phi )$ is slowly varying, we can ignore it's derivatives in the following calculations. Using the functional form of $g(\phi )$, the Hubble rate is equal to,
\begin{equation}\label{r11a}
H(t)=\frac{h(t)}{t}+h'(t)\ln\left (\frac{t}{t_0} \right ),
\end{equation}
and owing to the fact that the function $h(t)$ is slowly varying, the derivative $h'(t)$ can be safely ignored, and thereby the Hubble rate simplifies to,
\begin{equation}\label{r11b}
H(t)\simeq \frac{h(t)}{t},
\end{equation}
which is exactly equal to the LQC-radiation phase Hubble rate given in Eq. (\ref{r6}). This is a crucial point in our analysis, and the validity of the method is verified by exactly this coincidence. By using Eq. (\ref{r11}) and ignoring the derivatives $h'(t),h''(t)$ the differential equation (\ref{r5}) can be cast in the following form,
\begin{align}\label{r12}
& 2\frac{\mathrm{d}^2P(\phi (t))}{\mathrm {d}t^2}-\frac{h(\phi )}{\phi }\frac{\mathrm{d}P(\phi (t))}{\mathrm{d}t}-\frac{2h(\phi )}{\phi^2} P(\phi (t))=0.
\end{align} 
In order to find the $P(\phi (R))$ function from the above equation, we must first find the exact functional dependence of $\phi$ as a function of the Ricci scalar, which can be easily done if we calculate the Ricci scalar $R$ by taking into account relations (\ref{ricciscal}) and (\ref{r6}) and (\ref{r7}). Combining these, the final result reads, 
\begin{equation}\label{r13}
R(\phi )\simeq \frac{6\left (-h(\phi )+2h(\phi )^2  \right )}{\phi^2}.
\end{equation}
Notice that we ignored the higher derivatives of $h(t)$ in order to obtain Eq. (\ref{r13}). Using the exact form of $h(\phi )$ given in Eq. (\ref{r7}) (recall that $\phi=t$), solving Eq. (\ref{r13}) with respect to $\phi^2$, yields the result,
\begin{align}\label{r15}
& \phi^2 =-\frac{2}{3 q}+\frac{2}{R}-\frac{h_f}{3 R}+\frac{2 h_f^2}{3 R}\\ \notag &
+ \Big{(}4\times 2^{1/3} q+\frac{12\times 2^{1/3} q^2}{R}+\frac{2^{1/3} R}{3}+\frac{2^{1/3}q h_f}{3}\\ \notag & -\frac{4\times 2^{1/3} q^2 h_f}{R}- \frac{8\times 2^{1/3} q h_f^2}{3}+\frac{25\times 2^{1/3} q^2 h_f^2}{3R}-\frac{4\times 2^{1/3} q^2 h_f^3}{3R}+4\times 2^{1/3} q^2 h_f^4+\frac{1}{3 2^{1/3} q^2 R}\Big{)} \\ \notag &
\times \frac{1}{\left ( \alpha_0+\alpha_1R+\alpha_2R^2+\alpha_3R^3 +\sqrt{\beta_2R^2+\beta_3 R^3+\beta_4 R^4+\beta_5R^5} \right )^{1/3}}.
\end{align}
For simplicity, the coefficients $\alpha_i,\beta_i$ are given in appendix A. Using Eq. (\ref{r15}), we can solve the differential equation (\ref{r12}) and express the solutions as functions of the Ricci scalar. Then, by using Eq. (\ref{r1}) we can have the reconstructed $F(R)$ gravity. The general solution to the differential equation (\ref{r12}) is \cite{importantpapers3},
\begin{align}\label{r16}
& P(\phi )=c_1\phi^{\frac{h(\phi )-1+\sqrt{h(\phi)^2+6h(\phi )+1}}{2}}+c_2\phi^{\frac{h(\phi )-1-\sqrt{h(\phi)^2+6h(\phi )+1}}{2}},
\end{align}
and by using this we may easily find $Q(\phi)$ by substituting (\ref{r11}) in equation (\ref{r5a}), in which case we obtain,
\begin{align}\label{r18}
& Q(\phi )=-6h(\phi )c_1\left (h(\phi )+\frac{h(\phi )-1+\sqrt{h(\phi)^2+6h(\phi )+1}}{2}\right )\phi^{\frac{h(\phi )-1+\sqrt{h(\phi)^2+6h(\phi )+1}}{2}-2}\\ \notag &
-6h(\phi )c_2\left (h(\phi )+\frac{h(\phi )-1-\sqrt{h(\phi)^2+6h(\phi )+1}}{2}\right )\phi^{\frac{h(\phi )-1-\sqrt{h(\phi)^2+6h(\phi )+1}}{2}-2}
\end{align}
Since we are interested in the large $R$ regime of our approximate method, in the following we shall examine the solutions we found in the large $R$ limit only. In fact, it is an exercise of academic interest only to examine the small curvature limit, since in the context of the cosmological scenario we are studying, this era is governed by the CDM fluid, so it is senseless to examine the low curvature limit.

\subsection{$F(R)$ Gravity in the Large $R$ Limit}

Having solution (\ref{r15}) at hand, we may easily obtain the large $R$ limit, which is,
\begin{align}\label{r16a}
& \phi^2\simeq -\frac{1}{3 q}+\frac{\mathcal{A}_1}{R},
\end{align}
so finally we get,
\begin{equation}\label{r17a}
\phi \sim \sqrt{\frac{R-\mathcal{A}_1}{3qR}}
\end{equation}
which holds true when $R>\mathcal{A}_1$, where we have set $\mathcal{A}_1$ to be equal to,
\begin{align}\label{r18a}
\mathcal{A}_1=\frac{4\times 2^{1/3} q+\frac{2^{1/3}q h_f}{3}- \frac{8\times 2^{1/3} q h_f^2}{3}+4\times 2^{1/3} q^2 h_f^4+(2-h_f+2h_f^2)a_3^{1/3}}{a_3^{1/3}}.
\end{align}
In addition, the parameter $\alpha_3$ can be found in appendix A. We can find the function $P(\phi (R))$ in the small $\phi$ limit (or large $R$ limit equivalently), by recalling that the function $h(t)$ has the following limiting value as $\phi$ approaches zero,
\begin{equation}\label{r19}
\lim_{t\rightarrow 0}h(t)=0.
\end{equation} 
This observation significantly simplifies the calculation, since in that limit $P(\phi )$ reads,
\begin{equation}\label{r20}
P(\phi )=c_1+c_2\phi^{-1}.
\end{equation}
In virtue of Eq. (\ref{r17a}), the term $\phi^{-1}$ is equal to,
\begin{equation}\label{r17aevanescnewsols}
\phi^{-1} \sim \sqrt{\frac{3qR}{R-\mathcal{A}_1}}
\end{equation}
so eventually, the function $P(\phi )$ of Eq. (\ref{r20}) is equal to,
\begin{equation}\label{r20newexpr}
P(\phi )=c_1+c_2\sqrt{\frac{3qR}{R-\mathcal{A}_1}}.
\end{equation}
In the same vain, the function $Q(\phi (R))$ is found to be approximately equal to zero,
\begin{equation}\label{r23}
Q(\phi(R))\simeq 0.
\end{equation}
Combining Eqs. (\ref{r20newexpr}) and (\ref{r23}), the final form of the reconstructed $F(R)$ gravity of Eq. (\ref{r1}) reads,
\begin{equation}\label{r24newexpres}
F(R)\simeq c_1 R+c_2\sqrt{\frac{3qR^3}{R-\mathcal{A}_1}}
\end{equation}
We can further simplify the resulting $F(R)$ gravity, by expanding the square root in the large $R$ limit, in which case we get,
\begin{equation}\label{r24039494}
F(R)\simeq c_1 R+\frac{c_2\sqrt{3q}\mathcal{A}_1}{2}+c_2\sqrt{3q}R.
\end{equation}
which is rewritten,
\begin{equation}\label{r24}
F(R)\simeq \left ( c_1+c_2\sqrt{3q}\right ) R+\frac{c_2\sqrt{3q}\mathcal{A}_1}{2}.
\end{equation}
We can choose the coefficient of $R$ in the above expression to be equal to one, that is,
\begin{equation}\label{r24a}
c_1+c_2\sqrt{3q}=1,{\,}{\,}{\,}
\end{equation}
and defining $\Lambda$ in the following way,
\begin{equation}\label{r24b}
\Lambda =\frac{c_2\sqrt{3q}\mathcal{A}_1}{2},
\end{equation}
we end up to the final form of the reconstructed $F(R)$ gravity,
\begin{equation}\label{r24c}
F(R)\simeq R+\Lambda,
\end{equation}
which is Einstein-Hilbert gravity plus cosmological constant. This is a quite intriguing result, since the cosmological constant can describe inflationary dynamics of an expanding universe. Let us here recapitulate what we found in this section. We investigated which $F(R)$ gravity can generate the large curvature LQC-radiation cosmological scenario, which corresponds to the scale factor (\ref{holcor1}) and we ended up to an Einstein-Hilbert gravity plus a cosmological constant. Notice that the value of cosmological constant $\Lambda$ can be of the Planck energy order, since the parameter $q$ is related to the Planck energy-density (see Eq. (\ref{r7a})). Combining this result with the one we obtained in the $\Lambda$CDM study, which resulted to an $R^2$ gravity we have the physically appealing picture in which, when the large curvature limit of the two cosmological scenarios is considered, both reconstructed $F(R)$ gravity result to gravities that actually can generate inflationary solutions. And most importantly, in our case no matter fluids are present, so the inflationary dynamics is a result of pure $F(R)$ gravity. We shall further discuss this result in a later section.

\section{Stability of $F(R)$ gravity describing the $\Lambda$CDM bounce and the radiation bounce}

Before proceeding to some phenomenological applications of the modified gravity description for the $\Lambda$CDM bounce we provided, it is of critical importance to check the stability of our solutions. In this section we address this issue using the formalism developed in Ref. \cite{recontechniques}. We start off with the $F(R)$ gravity given in Eq. (\ref{frgenerlargetssss}), which describes the $\Lambda$CDM phase of the bounce solution.

\subsection{Study of the $F(R)$ stability for the matter-CC phase}

For the modified gravity of Eq. (\ref{frgenerlargetssss}) we used the reconstruction method developed in \cite{importantpapers3}, for which, the most appropriate stability check method is the one firstly done in \cite{recontechniques}, which we employ in this section. We perform a perturbation of the form,
\begin{equation}\label{pert1}
G(N)=g(N)+\delta g(N),
\end{equation}
and we insert this in Eq. (\ref{newfrw1modfrom}) noticing that $g(N)$ satisfies equation (\ref{newfrw1modfrom}). Consequently, the perturbation $\delta g(N)$ satisfies the following equation,
\begin{align}\label{stabpert1}
& g(N)\frac{\mathrm{d}^2F(R)}{\mathrm{d}R^2}\Big{|}_{R=3g'(N)+12g(N)}\delta ''g(N)
+\Big{[}3g(N)\left (4g'(N)+g''(N)\right )\frac{\mathrm{d}^3F(R)}{\mathrm{d}R^3}\Big{|}_{R=3g'(N)+12g(N)}
\\ \notag & +\left (3g(N)-\frac{1}{2}g'(N)\right )\frac{\mathrm{d}^2F(R)}{\mathrm{d}R^2}\Big{|}_{R=3g'(N)+12g(N)}\Big{]}\delta 'g(N)\\ \notag &
+\Big{[} 12g(N)\left(4g'(N)+g''(N)\right)\frac{\mathrm{d}^3F(R)}{\mathrm{d}R^3}\Big{|}_{R=3g'(N)+12g(N)}
\\ \notag &+\left(-4g(N)+2g'(N)+g''(N)\right)\frac{\mathrm{d}^2F(R)}{\mathrm{d}R^2}\Big{|}_{R=3g'(N)+12g(N)}+\frac{1}{3}\frac{\mathrm{d}F(R)}{\mathrm{d}R}\Big{|}_{R=3g'(N)+12g(N)}\Big{]}\delta g(N)=0.
\end{align}
Using this equation, the stability of the solution under small perturbations of $G(N)$ may be directly obtained. Indeed, the stability conditions read,
\begin{equation}\label{st0}
J_1=\frac{6(4 g'(N)+g''(N))F'''(R)}{F''(R)}+6-\frac{g'(N)}{g(N)}>0,
\end{equation}
and in addition, 
\begin{equation}\label{st01}
J_2=\frac{36 (4 g'(N)+g''(N)) F'''(R)}{F''(R)}-12+\frac{6 g'(N)}{g(N)}+\frac{3 g''(N)}{g(N)}+\frac{ F'(R)}{g(N) F''(R)}>0.
\end{equation}
From Eqs. (\ref{stabpert1}), (\ref{st0}), (\ref{st01}) and in conjunction with Eq. (\ref{newfrw1modfrom}), we can check the stability of the solution (\ref{frgenerlargetssss}), under the small perturbation (\ref{pert1}), in which case the stability conditions become, 
\begin{align}\label{stab1}
& J_1=6+3 (1-\delta)-2 3 (1-\delta) \Big{(}3^{\rho_1} c_1 \Big{(}-A \Big{(}-4+3 (1-\delta)\Big{)} e^{-3 (1-\delta) N}\Big{)}^{\rho_1} (-2+\rho_1) (-1+\rho_1) \rho_1 
\\ \notag & +3^{\rho_2} c_2 \Big{(}-A \Big{(}-4+3 (1-\delta)\Big{)} e^{-3 (1-\delta) N}\Big{)}^{\rho_2} (-2+\rho_2) (-1+\rho_2) \rho_2\Big{)} \\ \notag &
\times 3^{\rho_1} c_1 \Big{(}-A \Big{(}-4+3 (1-\delta)\Big{)} e^{-3 (1-\delta) N}\Big{)}^{\rho_1} (-1+\rho_1) \rho_1 
\\ \notag & +3^{\rho_2} c_2 \Big{(}-A \Big{(}-4 +3 (1-\delta)\Big{)} e^{-3 (1-\delta) N}\Big{)}^{\rho_2} (-1+\rho_2) \rho_2>0,
\end{align}
and in addition, the parameter $J_2$ reads,
\begin{align}\label{st2}
& J_2=-12-6 3 (1-\delta)+27(1-\delta)^2+e^{3 (1-\delta) N} \Big{(}c_1 \Big{(}12 A e^{-3 (1-\delta) N}-3 A 3 (1-\delta) e^{-3 (1-\delta) N}\Big{)}^{-1+\rho_1} \rho_1
\\ \notag &+c_2 \Big{(}12 A e^{-3 (1-\delta) N}-3 A 3 (1-\delta) e^{-3 (1-\delta) N}\Big{)}^{-1+\rho_2} \rho_2\Big{)}A^{-1} Q_1^{-1}
\\ \notag & +36 \Big{(}-4 A 3 (1-\delta) e^{-3 (1-\delta) N}+A c^4 e^{-3 (1-\delta) N}\Big{)} 
\\ \notag & \times \Big{(}c_1 \Big{(}12 A e^{-3 (1-\delta) N}-3 A 3 (1-\delta) e^{-3 (1-\delta) N}\Big{)}^{-3+\rho_1} (-2+\rho_1) (-1+\rho_1) \rho_1
\\ \notag & +c_2 \Big{(}12 A e^{-3 (1-\delta) N}-3 A 3 (1-\delta) e^{-3 (1-\delta) N}\Big{)}^{-3+\rho_2} (-2+\rho_2) (-1+\rho_2) \rho_2\Big{)}\times Q_1^{-1}>0,
\end{align}
where $Q_1$ is equal to,
\begin{align}\label{djhfj}
& Q_1=c_1 \Big{(}12 A e^{-3 (1-\delta) N}-3 A 3 (1-\delta) e^{-3 (1-\delta) N}\Big{)}^{-2+\rho_1} (-1+\rho_1) \rho_1
\\ \notag & +c_2 \Big{(}12 A e^{-3 (1-\delta) N}-3 A 3 (1-\delta) e^{-3 (1-\delta) N}\Big{)}^{-2+\rho_2} (-1+\rho_2) \rho_2,
\end{align}
and also $A=\frac{4 \text{a0}^{-3 (1-\delta )}}{9 (1-\delta )^2}$. Having the stability parameters $J_1$ and $J_2$ we can directly check the stability conditions for the most interesting values of $\delta$ we found in the previous sections. Particularly, for $\delta=1$, in which case the $F(R)$ gravity becomes $R^2$, and for $\delta=0.95$, $\delta=0.97$, for which values, the corresponding Einstein frame theory produces inflationary parameters which have exact concordance with the observational data. For $\delta=1$, the stability parameters become,
\begin{equation}\label{stpard1}
J_1=2>0,{\,}{\,}{\,}J_2=0,
\end{equation}
thus the $R^2$ does not satisfy the second stability solution, a fact that shows that this solution is unstable under perturbations. However, when $\delta$ takes the value $\delta\simeq 2$ but not exactly two, stability is ensured. In the same way, the stability parameters for $\delta=0.97$ become,
\begin{equation}\label{stpard2}
J_1=3 \left(2.03667\, +\frac{0.803333}{1+\frac{2.8063\times 10^{48} \text{c1} \left(\frac{e^{-0.09 N}}{a_0^{0.09}}\right)^{13.3889}}{\text{c2}}}\right)>0,
\end{equation}
with regards to $J_1$ which is obviously positive. As for $J_2$, this reads,
\begin{align}\label{sjhdghucdfvj}
& J_2=\frac{e^{-0.09N} \Big{(}4.71 10^{-49}c_2 \Big{(}\frac{e^{-0.09 N}}{a_0^{0.09}}\Big{)}^{27.2}-7.947c_1 \Big{(}\frac{e^{-0.09 N}}{a_0^{0.09}}\Big{)}^{40.61}\Big{)}}
{a_0^{0.09}\Big{(}2.2810^{-49} c_2\Big{(}\frac{e^{-0.09 N}}{a_0^{0.09}}\Big{(}2.2810^{-49}c_2\Big{(}\frac{e^{-0.09 N}}{a_0^{0.09}}\Big{)}^{28.2}+0.64 c_1\Big{(}\frac{e^{-0.09 N}}{a_0^{0.09}}\Big{)}^{41.61}\Big{)}}
\\ \notag &+
\frac{a_0{}^{0.18} e^{0.18 N} \left(-2.14 10^{-49}c_2 \left(\frac{e^{-0.09 N}}{a_0^{0.09}}\right)^{29.2}+8.4 c_1 \left(\frac{e^{-0.09 N}}{a_0{}^{0.09}}\right){}^{42.61}\right)}
{a_0^{0.09}\Big{(}2.2810^{-49} c_2\Big{(}\frac{e^{-0.09 N}}{a_0^{0.09}}\Big{(}2.2810^{-49}c_2\Big{(}\frac{e^{-0.09 N}}{a_0^{0.09}}\Big{)}^{28.2}+0.64 c_1\Big{(}\frac{e^{-0.09 N}}{a_0^{0.09}}\Big{)}^{41.61}\Big{)}},
\end{align}
with the most dominant term for any value of the e-folds number $N$ being,
\begin{equation}\label{mostdominant}
J_2\simeq \frac{a_0{}^{0.18} e^{0.18 N} \left(8.4 c_1 \left(\frac{e^{-0.09 N}}{a_0{}^{0.09}}\right){}^{42.61}\right)}
{a_0^{0.09}\Big{(}2.2810^{-49} c_2\Big{(}\frac{e^{-0.09 N}}{a_0^{0.09}}\Big{(}2.2810^{-49}c_2\Big{(}\frac{e^{-0.09 N}}{a_0^{0.09}}\Big{)}^{28.2}+0.64 c_1\Big{(}\frac{e^{-0.09 N}}{a_0^{0.09}}\Big{)}^{41.61}\Big{)}}>0
\end{equation}
which is obviously positive. Therefore the solution for $\delta=0.97$ is perfectly stable. The same applies for $\delta=0.95$, in which case we have,
\begin{equation}\label{evansed}
J_1=3 \left(2.03667\, +\frac{0.803333}{1+\frac{2.8063\times 10^{48} c_1 \left(\frac{e^{-0.09 N}}{a_0^{0.09}}\right)^{13.3889}}{c_2}}\right)>0,
\end{equation}
which is positive, as for $J_2$ the leading order contribution reads,
\begin{equation}\label{mostdominantqwd}
J_2\simeq \frac{8.46 a_0^{0.18} e^{0.18 N}c_1 \left(\frac{e^{-0.090 N}}{a_0{}^{0.09}}\right){}^{42.61}}
{a_0^{0.09}\Big{(}2.2810^{-49} c_2\Big{(}\frac{e^{-0.09 N}}{a_0^{0.09}}\Big{(}2.2810^{-49}c_2\Big{(}\frac{e^{-0.09 N}}{a_0^{0.09}}\Big{)}^{28.2}+0.64 c_1\Big{(}\frac{e^{-0.09 N}}{a_0^{0.09}}\Big{)}^{41.61}\Big{)}}>0,
\end{equation}
which is also positive. Therefore, as can be easily checked, for all values of $\delta $, except for $\delta=2$, stability is ensured. In addition, from an observational point of view phenomenologically most interesting cases $\delta=0.95$ and $\delta=0.97$, provide stable $F(R)$ solutions.

\subsection{Study of the $F(R)$ stability for the radiation bounce case}

We now examine the stability of the LQC-radiation generating $F(R)$ gravity given in Eq. (\ref{r24c}). Since this $F(R)$ gravity is the result of a reconstruction method, different to the previously used, we shall use the stability method developed in \cite{recontechniques}. We start from the following equation,
\begin{align}\label{staux1}
& 2\frac{\mathrm{d}^2P(\phi (t))}{\mathrm {d}t^2}-2g'(\phi )\frac{\mathrm{d}P(\phi (t))}{\mathrm{d}t}+4g''(\phi ) P(\phi (t))=0,
\end{align}
which can be rewritten in the following way,
\begin{align}\label{stauxo1}
&2\frac{\mathrm{d}^2P(\phi )}{\mathrm {d}\phi^2}\Big{(}\frac{\mathrm{d}\phi }{\mathrm {d}t}\Big{)}^2-2\frac{\mathrm{d}P(\phi )}{\mathrm {d}\phi }\frac{\mathrm{d}^2\phi }{\mathrm {d}t^2}-2g'(\phi )\frac{\mathrm{d}P(\phi )}{\mathrm {d}\phi}\Big{(}\frac{\mathrm{d}\phi }{\mathrm {d}t}\Big{)}^2+4\Big{(}g''(\phi )\Big{(}\frac{\mathrm{d}\phi }{\mathrm {d}t}\Big{)}^2+g'(\phi )\frac{\mathrm{d}^2\phi }{\mathrm {d}t^2}\Big{)}P(\phi )=0,
\end{align}
and after some calculations, it can be cast as follows,
\begin{align}\label{staux2}
& 2\Big{[}\frac{\mathrm{d}^2P(\phi )}{\mathrm {d}\phi^2}-g'(\phi)\frac{\mathrm{d}P(\phi )}{\mathrm {d}\phi}+g''(\phi )P(\phi )\Big{]}\Big{(}\Big{(}\frac{\mathrm{d}\phi }{\mathrm {d}t}\Big{)}^2-1\Big{)}
+2\Big{(}\frac{\mathrm{d}P(\phi )}{\mathrm {d}\phi}+2g'(\phi)P(\phi)\Big{)}\frac{\mathrm{d}^2\phi }{\mathrm {d}t^2}=0.
\end{align}
We define the function $\delta $ to be equal to,
\begin{equation}\label{staux3}
\delta =\frac{\mathrm{d}\phi }{\mathrm {d}t}-1.
\end{equation}
This parameter $\delta$ represents actually the way that perturbations behave for the solutions we found, since it practically measures the deviation of the parameter $\phi$ from $t$, which we considered to be identical. Using the parameter $\delta $, we can recast Eq.(\ref{staux2}) in the following way,
\begin{equation}\label{staux4}
\frac{\mathrm{d}\delta }{\mathrm{d}t}=-\omega (t)\delta,
\end{equation}
where $\omega (t)$ stands for,
\begin{equation}\label{omegarepr}
\omega (t)=2\frac{\frac{\mathrm{d}^2P(\phi )}{\mathrm {d}\phi^2}-g'(\phi)\frac{\mathrm{d}P(\phi )}{\mathrm {d}\phi}+g''(\phi )P(\phi )}{\frac{\mathrm{d}P(\phi )}{\mathrm {d}\phi}+2g'(\phi )P(\phi )}\Big{|}_{\phi=t}.
\end{equation}
If $\omega>0$ for the solution $P(\phi)$ we found then stability is ensured, since the perturbation becomes small. In the contrary case, instability occurs, since the perturbations are large. We shall examine the function $P(\phi)$ that corresponds to the large curvature limit, since this is the most interesting case. For the function $g(\phi )$ given in relation (\ref{r11}), $\omega $ becomes,
\begin{align}\label{hbdsbgd}
\omega (\phi )=\frac{-P(\phi ) \left(\frac{h(\phi )}{\phi }+\ln(\phi ) h'(\phi )\right)+2 P(\phi ) \left(-\frac{h(\phi )}{\phi ^2}+\frac{2 h'(\phi )}{\phi }+\ln(\phi ) h''(\phi )\right)+P''(\phi )}{2 P(\phi ) \left(\frac{h(\phi )}{\phi }+\ln(\phi ) h'(\phi )\right)+P'(\phi )},
\end{align}
and by neglecting the higher derivatives of $h(\phi)$, $\omega$ becomes,
\begin{align}\label{hbdsbgdsdfsf12345}
\omega (\phi )\simeq \frac{-P(\phi ) \left(\frac{h(\phi )}{\phi }\right)+2 P(\phi ) \left(-\frac{h(\phi )}{\phi ^2}\right)+P''(\phi )}{2 P(\phi ) \left(\frac{h(\phi )}{\phi }\right)+P'(\phi )}.
\end{align}
In this case, by using $P(\phi)$ as given in Eq. (\ref{r20}), the form of $h(\phi)$ given in (\ref{r7}) and substituting in Eq. (\ref{omegarepr}), we obtain,
\begin{align}\label{parameterpmega}
& \omega (\phi)\simeq \frac{2 c_2}{\phi ^3 \left(-\frac{c_2}{\phi ^2}+\frac{2 q \phi  \left(c_1+\frac{c_2}{\phi }\right) h_f}{1+q \phi ^2}\right)}-\frac{2 q c_1 h_f}{\left(1+q \phi ^2\right) \left(-\frac{c_2}{\phi ^2}
+\frac{2 q \phi  \left(c_1+\frac{c_2}{\phi }\right) h_f}{1+q \phi ^2}\right)}
\\ \notag & -\frac{q \phi  c_1 h_f}{\left(1+q \phi ^2\right) \left(-\frac{c_2}{\phi ^2}
 +\frac{2 q \phi  \left(c_1+\frac{c_2}{\phi }\right) h_f}{1+q \phi ^2}\right)}-\frac{q c_2 h_f}{\left(1+q \phi ^2\right) \left(-\frac{c_2}{\phi ^2}
+\frac{2 q \phi  \left(c_1+\frac{c_2}{\phi }\right) h_f}{1+q \phi ^2}\right)}
\\ \notag & -\frac{2 q c_2 h_f}{\phi  \left(1+q \phi ^2\right) \left(-\frac{c_2}{\phi ^2}+\frac{2 q \phi  \left(c_1+\frac{c_2}{\phi }\right) h_f}{1+q \phi ^2}\right)},
\end{align}
which for small $\phi$ (which corresponds to large curvatures) becomes,
\begin{equation}\label{lastapprox}
\omega (\phi)\simeq -\frac{2}{\phi }+q \phi,
\end{equation}
where $q$ is defined in Eq. (\ref{r7a}). Therefore $\omega$ is positive and thus stable for $q\phi^2>2$. On the contrary $\omega$ is negative for $q\phi^2<2$ and in this case the solution is unstable.

\section{Gravitational particle production}

As was also pointed out in Ref. \cite{lcdmbounce}, asymmetry of the bounce is required in order these cosmological bounce to be viable. One feature that can cause asymmetry of the bounce and that guarantees viability of the model, is particle production during the bounce. In this section we shall examine the gravitational particle production issue, adopting the research line of Refs. \cite{graviproduction}. As is well known \cite{graviproduction}, curvature oscillations can generate gravitational particle production. What mainly interests us is to examine the gravitational particle production for the matter-CC phase, which as we found is described by the $F(R)$ gravity given in Eq. (\ref{frgenerlargetssss}), since the LQC-radiation phase results to an $R+\Lambda$ gravity with known gravitational particle production properties \cite{birel}.

In the large curvature limit, the $F(R)$ gravity that generates the $\Lambda$CDM phase is an $R^2$ gravity, in which case there is sufficient particle production \cite{graviproduction} to guarantee the asymmetry of the bounce. We shall not go into details for this $F(R)$ gravity, since this issue was addressed in full detail in \cite{graviproduction}, but we shall be interested in the small curvature limit of the $\Lambda$CDM phase, in which case the $F(R)$ gravity is approximately equal to,
\begin{equation}\label{jfhfh}
F(R)\simeq c_2 R^{\rho_2}
\end{equation}
since $\rho_2$ is a negative number. Recalling the details with regards to the values that $\rho_2$ takes, since $0\leq \delta \leq 1$, the parameter $\rho_2$ takes the following values,
\begin{equation}\label{rho2paramevalues}
-\infty\leq \rho_2 <-0.128
\end{equation}
where the limiting values are obtained in the following limits,
\begin{equation}\label{limitiofrho}
\lim_{\delta\rightarrow 0}\rho_2=-0.128,{\,}{\,}{\,}\lim_{\delta\rightarrow 1}\rho_2=-\infty
\end{equation}
Let us briefly recall here how the $\Lambda$CDM bounce scenario works. At the beginning we have the bounce and when the curvature is quite large the radiation phase governs the dynamical evolution of the Universe. After this phase, the cosmological constant and cold dark matter governs the dynamics. At first, and when the curvature is too large, the cosmological constant phase occurs, which corresponds to $\delta=1$, and as the curvature lowers, the matter fluid governs the expansion. In the latter case, $\delta$ approaches zero. We shall study the gravitational production during that phase and for particular limits. Notice that in the end of the matter fluid matter-CC phase, a new bounce occurs, so this is a late time era, with time considered to start when the previous bounce had occurred in the past. Following \cite{graviproduction}, the equation that will reveal the particle production rate is the following,
\begin{equation}\label{generaleqngraviprod}
3 \square F_{,R}-R+RF_{,R}-2F(R)=0
\end{equation}
where we assumed that matter is absent. In Ref. \cite{graviproduction}, it was assumed that the spacetime is a flat Minkowski one, but here we assume a flat FRW metric of the form (\ref{metricformfrwhjkh}), in which case the Ricci scalar depends only on time. In this case, Eq. (\ref{generaleqngraviprod}) is simplified and by introducing the new variable $y=R^{\rho_2-1}$ and also using the form of the $F(R)$ gravity given in Eq. (\ref{jfhfh}) , Eq. (\ref{generaleqngraviprod}) becomes,
\begin{equation}\label{generaleqngraviprodnew}
3 \rho_2\partial_t^2y+3H\partial_ty-y^{1/(\rho_2-1)}+(\rho_2-2)y^{\rho_2/(\rho_2-1)}=0
\end{equation}
Keeping only the dominant terms in the small curvature limit, Eq. (\ref{generaleqngraviprodnew}) gets simplified,
\begin{equation}\label{generaleqngraviprodnew}
3 \rho_2\partial_t^2y+3H\partial_ty+(\rho_2-2)y^{\rho_2/(\rho_2-1)}=0
\end{equation}
where $H$ is the Hubble rate (\ref{hubble1}). We shall study the above equation in various limits in order to see how particle production behaves in the small curvature limit. As was proven in \cite{graviproduction}, the rate of gravitational production is given by,
\begin{equation}\label{rateofgraviprod}
\dot{\rho}_{PP}\simeq \frac{\Delta_R\omega}{1152 \pi}
\end{equation}
where $\Delta_R$ is the amplitude of the curvature oscillations and $\omega$ their physical frequency. Note that $\Delta_R$ is a slowly varying function of $t$ in general. The method used in \cite{graviproduction} required that the solution $y$ is definitely written in the following form,
\begin{equation}\label{slowvarpapprox}
R(t)=y^{1/(\rho_2-1)}(t)=\Delta_R (t) \sin (\omega t)
\end{equation}
since a general analytic solution can be quite difficult. In our case we shall find analytic solutions of (\ref{generaleqngraviprodnew}), with one of them being exactly of the form (\ref{slowvarpapprox}), without any approximation. The other solution we shall present is analytic but we fit the resulting function in such a way that we end up to a function of the form (\ref{slowvarpapprox}). We start our investigation with the assumption that $\delta\rightarrow 1$ and also for times such that,
\begin{equation}\label{firstlim}
\frac{2\rho_2}{1-\delta}\gg t
\end{equation}
Practically, this means that we are dealing with times after the LQC-radiation phase, and during the start of the $\Lambda$CDM phase. In this case, by keeping the dominant terms in Eq. (\ref{rateofgraviprod}), the latter becomes,
\begin{equation}\label{latterbehav}
3H\partial_ty+(\rho_2-2)y^{\rho_2/(\rho_2-1)}=0
\end{equation}
which has a simple solution,
\begin{equation}\label{soly1}
y(t)=ce^{\frac{-(t-c_1)(1-\delta)}{2}}
\end{equation}
The function (\ref{soly1}) if written in the form (\ref{slowvarpapprox}), which can be done by a numerical fit, yields the following rate of gravitational particle production,
\begin{equation}\label{rateofgraviprodena}
\dot{\rho}_{PP}\simeq 6079
\end{equation}
where we used $\delta=0.99$ for which case $\rho_2\simeq-33.19$. Notice that this rate strongly depends on the fraction $(1-\delta)/\rho_2$ and on the arbitrary parameter $c$. We proceed to another interesting limiting case, which has an interesting analytic solution. Particularly, we are interested in the late time evolution, and particularly for the cosmological time being,
\begin{equation}\label{firstlimthirdlimit}
t\gg \frac{2\rho_2}{1-\delta}
\end{equation}
In this case, and as $\delta\rightarrow 1$, the differential equation (\ref{generaleqngraviprodnew}) takes the following form,
\begin{equation}\label{generaleqngraviprodnewlev}
3 \rho_2\partial_t^2y+(\rho_2-2)y^{\rho_2/(\rho_2-1)}=0
\end{equation}
which is the Emden-Fowler differential equation. Setting,
\begin{equation}\label{setlocal}
m=\frac{\rho_2}{\rho_2-1},{\,}{\,}{\,}A=-\frac{\rho_2-2}{3\rho_2}
\end{equation}
the solution of (\ref{generaleqngraviprodnewlev}) for $m\neq -1$ becomes,
\begin{equation}\label{gensolemdenfowl}
t=yF_1(\frac{1}{2},\frac{1}{1+m},1+\frac{1}{1+m},-\frac{2A}{C_a(1+m)}y^{1+m})+C_b
\end{equation}
with $F_1$ being the Gauss hypergeometric function and $C_a$, $C_b$ arbitrary integration constants. When $\delta\rightarrow 1$, the parameter $m$ is approximately equal to $m\simeq 1$, since only $\rho_2$ dominates. In addition, $A=-1/3$ therefore, in this limit, the solution (\ref{gensolemdenfowl}) is written,
\begin{equation}\label{gensolemdenfowl11}
t=yF_1(\frac{1}{2},\frac{1}{2},1+\frac{1}{2},\frac{1}{3 C_a }y^2)+C_b
\end{equation}
and by making the replacement $z=\sqrt{\frac{1}{3 c_a}}y$, the Gauss hypergeometric function $F_1$ has the following functional form,
\begin{equation}\label{gausshyperform}
F_1(\frac{1}{2},\frac{1}{2},1+\frac{1}{2},z^2)=z^{-1} \arcsin (z)
\end{equation} 
Using (\ref{gausshyperform}), solution (\ref{gensolemdenfowl11}) in terms of $y(t)$ becomes, 
\begin{equation}\label{gensolemdenfowl11hdfgyfdg}
t= \sqrt{3 c_a} \arcsin (\frac{1}{3 C_a}y)+C_b=
\end{equation}
so finally $y(t)$ is exactly equal to,
\begin{equation}\label{fngravpartoscilasol}
y(t)=\frac{1}{3 C_a}\sin \left ( \frac{1}{3 C_a}(t-C_b)\right )
\end{equation}
Note that we arrived at the solution (\ref{fngravpartoscilasol}) using only analytic methods, without any approximations. Notice that $y(t)$ is exactly of the form (\ref{slowvarpapprox}), so in the present case, the amplitude and the frequency of the curvature oscillations are,
\begin{equation}\label{freqsquare}
\Delta_R=\frac{1}{3 C_a},{\,}{\,}{\,}\omega=\frac{1}{3 C_a}
\end{equation}
Consequently, the rate of the gravitational particle production is,
\begin{equation}\label{lastequatrateofgraviprod}
\dot{\rho}_{PP}\simeq \frac{1}{(3 C_a)^21152 \pi}
\end{equation}
Since the parameter $C_a$ is a free parameter of the theory, it can be chosen to be quite small, so that the particle production rate is as big it is required in order to obtain an asymmetric bounce. Therefore, in our $F(R)$ gravity description of the $\Lambda$CDM bounce, sufficient gravitational particle production is ensured to guarantee an asymmetric bounce, during of course the matter-CC phase we described above.

\section*{Discussion}

We provided a pure $F(R)$ gravity description of the $\Lambda$CDM bounce scenario that was developed in Ref. \cite{lcdmbounce}, without the need for perfect matter fluids to govern the Universe's dynamics. The $\Lambda$CDM bounce scenario consists of two distinct cosmological eras, namely the radiation dominated and the matter-CC phase, and is therefore based on the discontinuity of the equation of state between the two eras. Using very well known reconstruction techniques, we were able to find which pure $F(R)$ gravity can generate each cosmological era. In the case of the radiation phase, the curvature is considered to be large, so the scale factor is $a(t)\sim (at^2+1)^{1/4}$, a result that is obtained by using LQC considerations. In the large curvature regime, such a cosmological expansion is generated by an $F(R)\sim R+\Lambda$ gravity, plus non dominant curvature terms in this approximation. In addition, the matter-CC era is generated by a power law $F(R)$ gravity of the form $F(R)\sim c_1R^{\rho_1}+c_2R^{\rho}$, with $\rho_i$ being numbers related to the details of the scale factor. In the large curvature regime, this $F(R)$ gravity is described by an $R^2$ gravity, while as curvature lowers, the $F(R)$ gravity takes the form $R^{\rho_1}$ with $\rho_1<2$. In this high curvature regime we found two particular values of the parameter $\rho_1$, for which the $F(R)$ gravity, when studied in the Einstein frame, yields results that have concordance with the latest Planck data on inflation. 

Interestingly enough, in the context of $F(R)$ gravity we found a solution to a problem that the $\Lambda$CDM bounce model of \cite{lcdmbounce} was confronted with. Particularly, in order that the $\Lambda$CDM bounce model is considered viable, the bounce must be asymmetric. As was pointed out in \cite{lcdmbounce}, one way to achieve this is through particle production during the bounce and in the framework of $F(R)$ gravity this process occurs naturally. As we explicitly demonstrated, particle production is particularly enhanced during the matter-CC phase, thus the asymmetry of the bounce can be ensured with the $F(R)$ gravity description of the $\Lambda$CDM bounce. Note finally that using the same method one can reconstruct $\Lambda$CDM bounce universe in $F(G)$ and $F(T)$ gravity (see Ref. \cite{sergeibounce}) this will be done elsewhere. 

A very important remark is in order. We have to mention that is of fundamental importance to explain why nature should select the discontinuous change in the equation of state. This is closely connected to the LQC effects, which effectively modify the large curvature era and govern the early time evolution of the bounce. This discontinuity of the equation of state appears in the $F(R)$ gravity description we provided, since in the large curvature regime, the LQC radiation phase is generated by an almost Einstein-Hilbert gravity, while the large curvature regime of the matter-CC phase is governed by an $R^2$. It would be quite interesting to find a natural explanation of this discontinuity in the context of Jordan frame $F(R)$ gravity, an issue we hope to address in a future publication.

It is important to discuss the possibility that the $F(R)$ realizations of the $\Lambda$CDM bounce we just presented, can be distinguished from other theoretical implements. In the case of Ref. \cite{lcdmbounce}, it was explicitly demonstrated that the running of the spectral index is negative, and therefore it can be tested from CMB observations. In the case of $F(R)$ gravity now, in principle a difference may come from the calculation of the spectral index, since it is calculated in a different way. But before getting into the details of this, let us mention that prior of distinguishing $F(R)$ gravity prediction, it is important to find seeds of a bouncing cosmology to the observational data, since there exist many bouncing scenarios. In addition, a quite interesting scenario was presented in Refs. \cite{piaocmb}, where a contracting bouncing phase preceded the slow-roll inflationary phase. With respect to the latter scenario, this could have observable effects in the CMB anomalies. By determining the exact evolutionary scenario, in the context of $F(R)$ gravity, the observational indices can be calculated explicitly in two frames, namely in the Jordan frame \cite{jordanframesergei} and in the Einstein frame \cite{sergeistarobinsky}. The Einstein frame calculation involves a scalar field on which the slow-roll conditions are imposed. Interestingly enough, when the calculation is performed in the Jordan frame, it is possible to have results that can directly be fitted to the observational data \cite{jordanframesergei}, without the need for imposing the slow-roll conditions. Indeed, in the Jordan frame, if no matter is present, as in our case, there is no scalar field and hence contact with observations can be done by using the technique of maximum likelihood. In this way, the parameters of the theory can be appropriately adjusted, so that the corresponding observational indices are produced. The observational indices have quite complicated form to be explicitly presented here, but can be found in the Appendix C of Ref. \cite{jordanframesergei}. Notice that pure $F(R)$ gravity models in the Jordan frame are in principle less restricted from the corresponding Einstein frame counterparts, owing to the fact that the parameters can be appropriately chosen. Hence, in order to distinguish these from other theoretical descriptions, other observational quantities must be examined, in addition to the aforementioned observables, where the differences could be significant, like for example in the growth index or similar quantities. Another possibility to distinguish predictions of modified gravity from General Relativity, may be related with the description of compact massive (neutron) stars but this goes beyond this work.

Finally, of the most sound results obtained in Re. \cite{lcdmbounce} is the existence of red tilted long wavelength perturbation modes. It would be therefore very important to study the long wavelength perturbations in the context of Jordan frame $F(R)$ theories. A study of perturbations valid to a certain limit was done in \cite{perturbations}. Of course the Einstein frame analysis can be dealt with standard techniques, but if someone addresses the full problem in the Jordan frame, this study can be quite difficult. We hope to address these issues in a future work.

\section*{Acknowledgments}

The research by S.D.O has been supported by MINECO (Spain) project FIS2010-15640 and FIS2013-44881 and by the Russian Government Program of Competitive Growth of Kazan Federal University.

\section*{APPENDIX A: Detailed presentation of polynomial coefficients}

In this appendix shall give in detail the polynomial coefficients $\alpha_i,\beta_j$, with $i=0,..3$ and $j=0,...5$, which appear in Eq. (\ref{r15}). Particularly, the $\alpha_i$'s are the coefficients of the following polynomial:
\begin{equation}\label{a1}
\rho (R)=\alpha_0+\alpha_1R+\alpha_2R^2+\alpha_3R^3
\end{equation} 
while the $\beta_j$'s are the coefficients of,
\begin{equation}\label{a2}
P(R)= \beta_2R^2+\beta_3 R^3+\beta_4 R^4+\beta_5R^5
\end{equation}
These are given in detail below,
\begin{align}\label{a3}
& \alpha_0=432 q^6-216 q^6 h_f+468 q^6 h_f^2-146 q^6 h_f^3+156 q^6 h_f^4-24 q^6 h_f^5+16 q^6 h_f^6 \\ \notag &
\alpha_1=216 q^5-18 q^5 h_f-75 q^5 h_f^2+30 q^5 h_f^3-48 q^5 h_f^4 \\ \notag & 
\alpha_2= 36 q^4+3 q^4 h_f+30 q^4 h_f^2 \\ \notag &
\alpha_3= 2 q^3 \\ \notag &
\beta_2= 45684 q^{10} h_f^2-15228 q^{10} h_f^3+31725 q^{10} h_f^4-5076 q^{10} h_f^5+5076 q^{10} h_f^6 \\ \notag &
\beta_3 = 23004 q^9 h_f^2+594 q^9 h_f^3-12528 q^9 h_f^4-216 q^9 h_f^5 \\ \notag &
\beta_4 =3861 q^8 h_f^2+540 q^8 h_f^3-108 q^8 h_f^4 \\ \notag &
\beta_5 = 216 q^7 h_f^2
\end{align}

\end{document}